\documentclass[aps,showpacs,preprintnumbers,amsmath,nofootinbib]{revtex4}
\usepackage{amssymb}
\usepackage{graphics}
\usepackage{graphicx}
\usepackage{dcolumn}
\usepackage{bm}
\usepackage{color}

\begin{document}


\title{Analysis of $pp$ and $\bar{p}p$ Elastic Scattering Based on Theoretical Bounds in High-Energy Physics: an Update}

\author{V. A. Okorokov} \email{VAOkorokov@mephi.ru; Okorokov@bnl.gov}
\affiliation{National Research Nuclear University MEPhI (Moscow
Engineering Physics Institute), Kashirskoe highway 31, 115409
Moscow, Russia}

\author{S. D. Campos} \email{sergiodc@ufscar.br}
\affiliation{Departamento de F\'isica, Qu\'imica e Matem\'atica, Universidade Federal de S\~ao Carlos, 18052-780, Sorocaba, SP, Brazil}

\date{\today}

\begin{abstract}
In a previous work a novel parametrization was proposed for the $pp$ and $\bar{p}p$ total cross section. Here, results are presented for analysis updated with taken into account the recent data from accelerator experiments as well as from cosmic ray measurements. The analytic parameterizations suggested within axiomatic quantum field theory (AQFT) provide the quantitative description of energy dependence of global scattering observables with robust values of fit parameters. Based on the fit results the estimations are derived for the total cross section and the $\rho$ parameter in elastic $pp$ scattering at various $\sqrt{s}$ up to energy frontier $\sqrt{s}=10$ PeV which can be useful for present and future hadron colliders as well as for cosmic ray measurements at ultra-high energies.

\end{abstract}


\maketitle

\section{\label{intro}Introduction}
The high energies physics is plentiful of open problems, despite the incredible success of the field theoretic approach, named Quantum Chromodynamics (QCD), in solving problems on this subject. As is well known, the hadron-hadron interactions at short distances, can only be studied using the perturbative approach of QCD. However, interactions at large distances (near the forward scattering) cannot be evaluated even within perturbation techniques. Therefore, the only way at present day to study a global parameters for hadron-hadron scattering is the using model approaches based on both formal theoretical results and experimental available data. In general, this modeling technique allows the formulation of a very large number of models each one enlightening some aspects of the collision process. Of course, all these efforts are attempts to construct a complete theory describing the hadron-hadron elastic and inelastic scattering.

The absence of a pure QCD description allows, as is well known, phenomenological analysis based on models \cite{modelos} or almost-independent models \cite{sdc} whose main goal is to fit the available experimental data extracting novel information, contributing to the development of useful computational schemes. Moreover, analyticity and crossing-symmetry are crucial theoretic results, however, important asymptotic theorems and bounds have been demonstrated in the last decades within the AQFT as Froissart--Martin and Jin--Martin bounds, providing formal constraints in the region of high-energies, which cannot be disregarded in any reliable formalism, mainly related with model-independent approaches.

In the recent scenario \cite{bc,sdcvao_2}, the introduction of a fractal dimension and phase transition point of view can be used for the total
cross section description as novel interpretation of the results discussed in \cite{sdcvao} and those obtained here.
Moreover, the results described in \cite{antoniou_5} must also be taken into account in order to analyze the results obtained
in the present paper.

The paper is organized as follows. In section \ref{sec2}, we present the basic formalism developed in \cite{sdcvao} for elastic $pp$ and $\bar{p}p$ scattering. Section \ref{sec3} contains the results for simultaneous fits of global scattering parameters with taking into
account the recent improvement for experimental database. Also the predictions for total cross sections in $pp$ and $\bar{p}p$
elastic scattering are briefly discussed for some energies. Section \ref{sec4} presents the final remarks and summary.

\section{\label{sec2}Global Parameters in Elastic Scattering}
For the sake of simplicity one neglects spin contributions as the energy increases, therefore the global physical quantities characterizing the elastic scattering are the well-known total cross section ($\sigma_{\footnotesize\mbox{tot}}$) and the ratio of the real to the imaginary part of the forward scattering amplitude ($\rho$ parameter). Therefore the set of global parameters $\mathcal{G}_{NN}=\{\sigma_{\scriptsize{\mbox{tot}}}^{pp},\sigma_{\scriptsize{\mbox{tot}}}^{\bar{p}p},\rho^{pp},\rho^{\bar{p}p}\}$ is under study for elastic $pp$ and $\bar{p}p$ scattering below.

The invariant amplitude $F(s,q^2)$ can be expressed as a function of two Mandelstam variables in the center-of-mass system, the
energy squared $s$ and the momentum transfer squared $-t=q^2$ GeV$^2$ as
\begin{eqnarray}
\nonumber F(s,q^2)=\mathrm{Re}F(s,q^2)+i\mathrm{Im}F(s,q^2),
\end{eqnarray}
\noindent where the imaginary part of $F$ represents the absorptive term in the scattering reaction. In a previous work \cite{sdcvao}, the scattering amplitude in forward direction, $-t=q^2=0$, was obtained based on both the formal high-energy bounds and the theorems from AQFT. Consequently, in \cite{sdcvao} the following parameterizations for the global scattering parameters were deduced within AQFT\footnote{In the equations below $s$ means dimensionless variable $s \equiv s/s_{0}$, where $s_{0}$=1 GeV$^{2}$, unless
otherwise specifically indicated.}
\begin{eqnarray}
\label{eq:2.1} \sigma_{\scriptsize{\mbox{tot}}}^{\pm}(s)&=&\frac{\delta^{\pm}}{s^{6+\gamma^{\pm}}} +
\frac{\pi\beta^{\pm}}{\sqrt{2}s^{\alpha^{\pm}}m_{\pi}^2} \ln^2s, \\
\label{eq:2.2} \rho^{\pm}(s)&=&\frac{1}{\sigma_{\scriptsize{\mbox{tot}}}^{\pm}}\biggl\{\frac{k}{s}+\frac{\pi}{2}\biggl[-\frac{\delta^{\pm}(6+\gamma^{\pm})}
{s^{6+\gamma^{\pm}}}+\frac{\pi\beta^{\pm}\sqrt{2}}{m_{\pi}^2s^{\alpha^{\pm}}}\biggl(\ln
s-\frac{\alpha^{\pm}\ln^2
s}{2}\biggr)\biggr] \pm \frac{\pi}{4}[\sigma_{\footnotesize\mbox{tot}}
^{+}-\sigma_{\footnotesize\mbox{tot}}^{-}]\biggr\}.
\end{eqnarray}

\noindent where the plus sign is for $pp$ and the minus one for $\bar{p}p$, the set of free parameters $\mathcal{P}_{NN} =
\{\alpha^{\pm},\beta^{\pm},\delta^{\pm},\gamma^{\pm},\kappa\}$ can be obtained from the fitting procedures of experimental data \cite{sdcvao}. Qualitatively speaking, the Froissart--Martin bound was never improved and, therefore, $\alpha^{+}$($\alpha^{-}$) may
be viewed as an empirical correction to this bound in order to tame its divergence as $s\rightarrow\infty$. The first term in the right-hand side of (\ref{eq:2.1}) is responsible for the experimental data description at low energies. Then, one can expect that $\delta^{+} \approx \delta^{-}$ and $\gamma^{+} \approx \gamma^{-}$, furthermore at very high energies $6+\gamma^{\pm} \to 0$. However, the second term in the right-hand side of (\ref{eq:2.1}) is the leading one at high-energies for both $pp$ and $\bar{p}p$ scatterings. Based on the Pomeranchuk theorem one expects, $\beta^{+} \to \beta^{-}$, $\alpha^{+} \to \alpha^{-}$ at very high energies and, usually, it is interpreted as a pomeron exchange.

\section{\label{sec3}Simultaneous Fits for Global Scattering Parameters}
It should be noted the database (DB) to set of the global scattering parameters $\mathcal{G}_{NN}$ was updated significantly since the previous analysis \cite{sdcvao} and the new experimental results were added for $\sigma_{\scriptsize{\mbox{tot}}}^{pp}$ mostly.

Our approach for phenomenological study of elastic $pp$ and $\bar{p}p$ scattering supposes the using of the newest available version of \textit{data files} of the Particle Data Group (PDG) for experimental results for global scattering parameters in order to avoid additional uncertainties and provide a correct comparison with other studies, in particular, previous analysis \cite{sdcvao}. But the latest available version of the PDG DB from \cite{pdg2016} does not contain the recent LHC results\footnote{Furthermore, the direct comparison made within the present work shows that DB16 from \cite{pdg2016} is identical to the previous version of the PDG DB -- DB14 from \cite{pdg2014} with regard of the set $\mathcal{G}_{NN}$ under consideration.} for global parameters in elastic $pp$ scattering at $\sqrt{s}=7$ TeV \cite{ATLAS-NPB-889-486-2014} and 8 TeV \cite{ATLAS-PLB-761-158-2016,TOTEM-EPJC-76-661-2016} as well as cosmic ray point at $\sqrt{s}=95^{+5}_{-8}$ TeV \cite{TA-PRD-92-032007-2015}. Therefore the simultaneous fit is made within the present paper for energy dependence of terms of the $\mathcal{G}_{NN}$ for different stages of the PDG DB: DB16, only the DB
from \cite{pdg2016}, and DB16+, the DB16 with new the LHC and cosmic ray data\footnote{It should be stressed the TOTEM results from \cite{TOTEM-NPB-899-527-2015} are not analyzed here because the latest results are available in \cite{TOTEM-EPJC-76-661-2016} for the same kinematic domain. It is fully within the rules were applied for creation of experimental databases in various analyzes, for instance, for elastic slope \cite{okorokov-arxiv-1501.01142} as well as in jet physics \cite{Okorokov-IJMPA-27-1250037-2012} and femtoscopy \cite{Okorokov-AHEP-2015-790646-2015}.}.

\begin{figure}
\centering
\includegraphics[width=17.0cm,height=17.0cm]{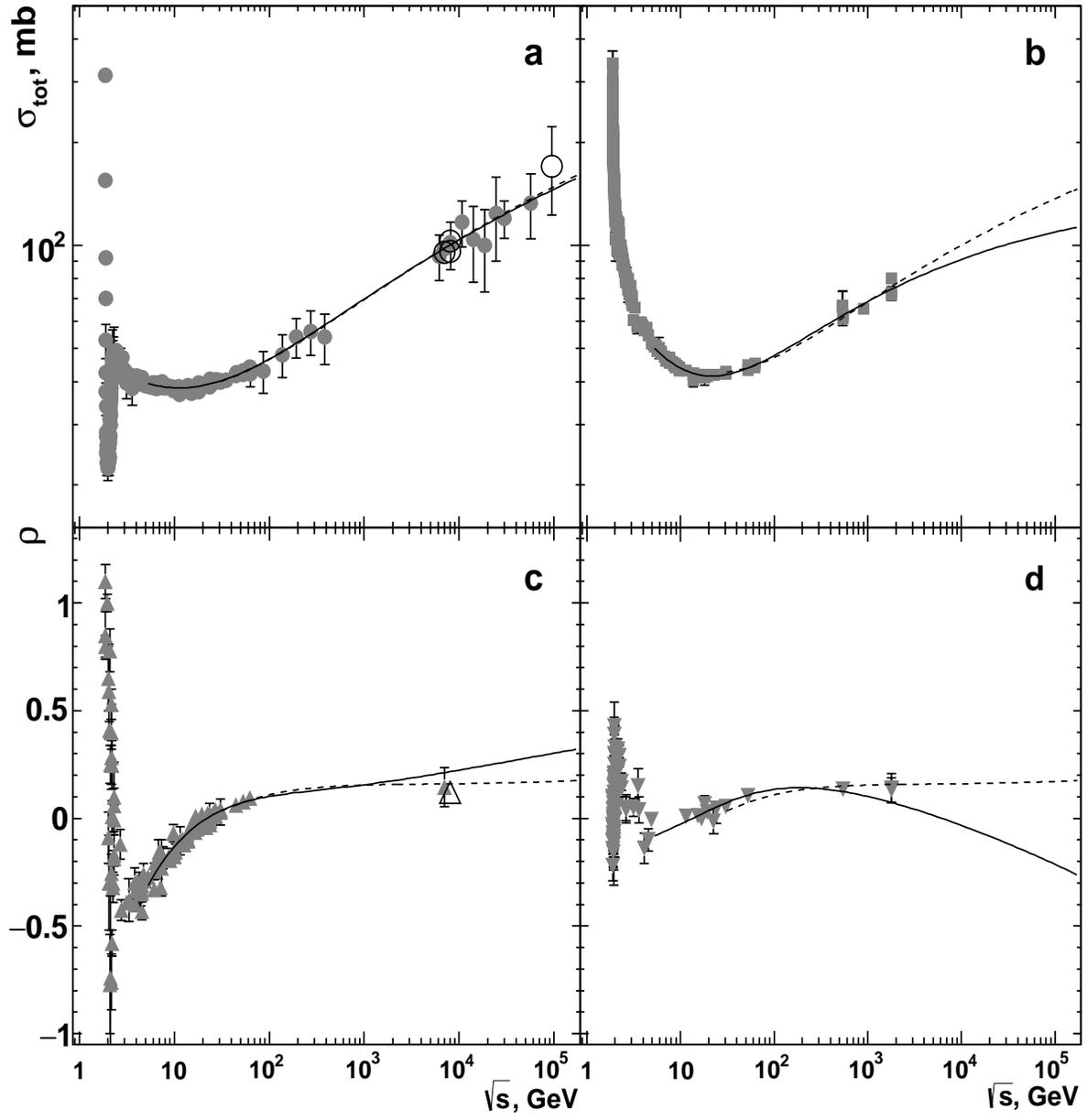}
\caption{The $\sigma_{\scriptsize{\mbox{tot}}}$ energy dependence for proton-proton
(a), antiproton-proton (b) scattering, $\rho^{pp}$ (c),
$\rho^{\bar{p}p}$ (d) parameters versus on collision energy and results
of simultaneous fits of all four parameters by
equations (\ref{eq:2.1}), (\ref{eq:2.2}). Solid points are from DB16 \cite{pdg2016} and these points are included in the fitted sample, open points from \cite{ATLAS-NPB-889-486-2014,ATLAS-PLB-761-158-2016,TOTEM-EPJC-76-661-2016,TA-PRD-92-032007-2015} are not included in the fitted data sample (detailed explanation -- see in text). The solid line
corresponds to the fit at $\sqrt{s_{\scriptsize{\mbox{min}}}}=5$ GeV, the dashed line
- at $\sqrt{s_{\scriptsize{\mbox{min}}}}=30$ GeV.} \label{fig:1}
\end{figure}

\begin{figure}
\centering
\includegraphics[width=17.0cm,height=17.0cm]{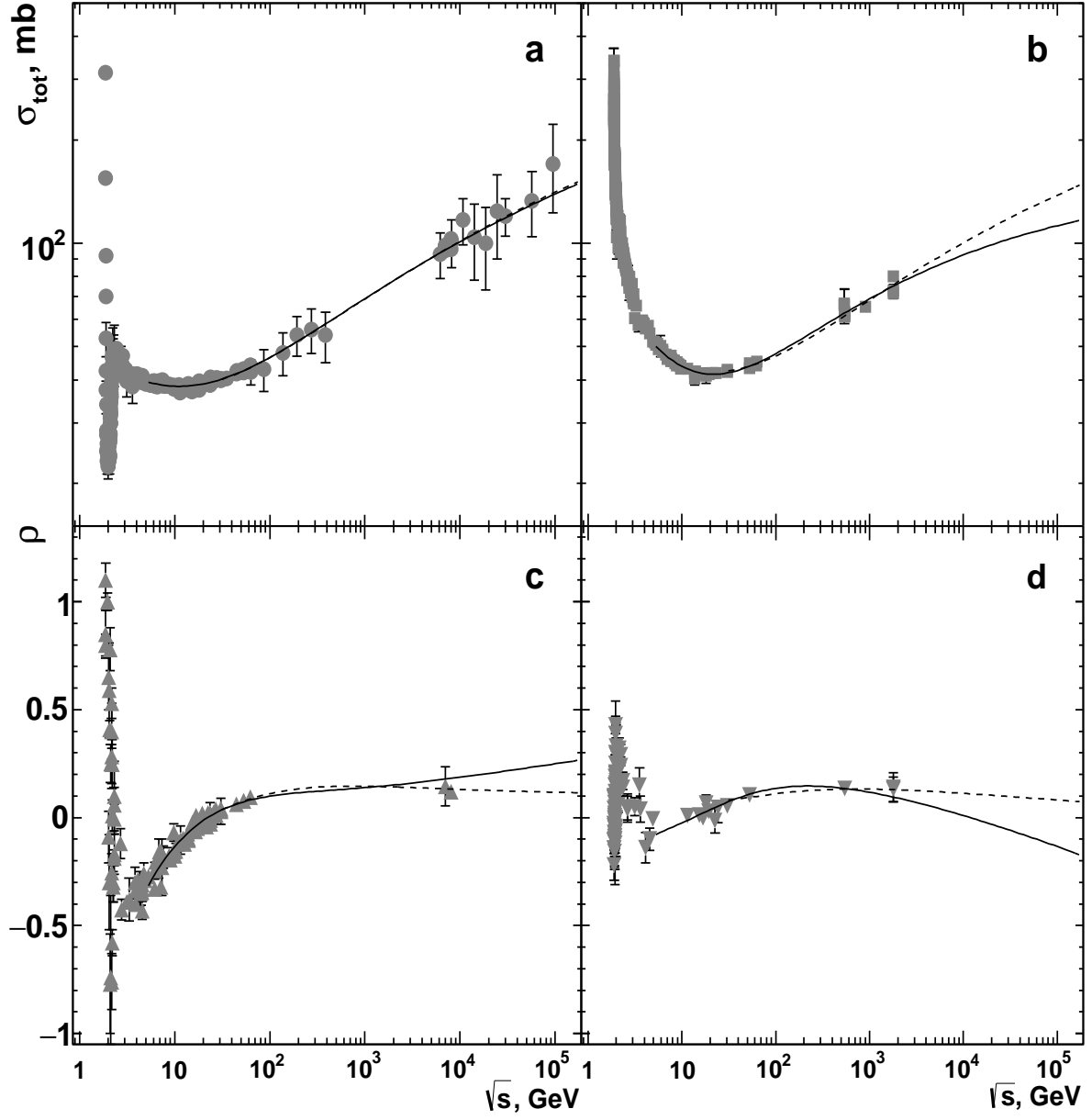}
\caption{The $\sigma_{\scriptsize{\mbox{tot}}}$ energy dependence for proton-proton
(a), antiproton-proton (b) scattering, $\rho^{pp}$ (c),
$\rho^{\bar{p}p}$ (d) parameters versus on collision energy and results
of simultaneous fits of all four parameters by
equations (\ref{eq:2.1}), (\ref{eq:2.2}). Experimental points from DB16+ \cite{pdg2016,ATLAS-NPB-889-486-2014,ATLAS-PLB-761-158-2016,TOTEM-EPJC-76-661-2016,TA-PRD-92-032007-2015} are included in the fitted sample. The solid line
corresponds to the fit at $\sqrt{s_{\scriptsize{\mbox{min}}}}=5$ GeV, the dashed line
- at $\sqrt{s_{\scriptsize{\mbox{min}}}}=30$ GeV.} \label{fig:2}
\end{figure}

Two methods are used for calculation of total errors, namely, (i) as linear sum of systematic and statistical errors and (ii) as addition of the two uncertainties in quadrature. The method (i) corresponds to the \cite{sdcvao,Avila-PRD-67-054020-2003} and allows the correct comparison with our previous analysis but the method (ii) is the same as well as in PDG approach \cite{pdg2016}. Simultaneous fits are made by equations (\ref{eq:2.1}), (\ref{eq:2.2}) for the energy dependence of global scattering parameters from the set $\mathcal{G}_{NN}$ for DB16 and DB16+ for each of the methods for total error calculations. The numerical results of the simultaneous fits are shown in Table \ref{table:3-1} and \ref{table:3-2} for method (i) and (ii) respectively. Experimental data with total errors from method (ii) together with fit curves are shown in Fig. \ref{fig:1} for DB16 and in Fig. \ref{fig:2} for DB16+. In the first case the new data with respect of the DB16 are shown also for completeness, but as indicated above the points from DB16 are only fitted in Fig. \ref{fig:1}. In Figs. \ref{fig:1}, \ref{fig:2} solid lines correspond to the fits at $\sqrt{s_{\scriptsize{\mbox{min}}}}=5$ GeV and dashed curves are fits at $\sqrt{s_{\scriptsize{\mbox{min}}}}=30$ GeV.

The values of all terms of the set $\mathcal{P}_{NN}$ from Table \ref{table:3-1} for DB16 coincide with the values of corresponding fit parameters from \cite{sdcvao} within errors\footnote{It should be noted that here $\beta^{\pm}$ parameters did not contain the conversation constant $(\hbar c)^{2}$ from \cite{pdg2016} in difference with the previous analysis \cite{sdcvao}.} for any $s_{\scriptsize{\mbox{min}}}$. As consequence all features of the fit curves in Fig. \ref{fig:1} agree with detailed discussion for previous analysis \cite{sdcvao}. It should be noted that fit qualities are better for the present study for DB16 than that for earlier simultaneous fit \cite{sdcvao} for an identical method of total error calculation at any $s_{\scriptsize{\mbox{min}}}$ and this improvement amplifies with the growth of the $s_{\scriptsize{\mbox{min}}}$.

As seen suggested approximation functions (\ref{eq:2.1}), (\ref{eq:2.2}) provide reasonable fit qualities at smallest $\sqrt{s_{\scriptsize{\mbox{min}}}}=3$ GeV and statistically acceptable $\chi^{2}/\mbox{n.d.f.}$ at any other $\sqrt{s_{\scriptsize{\mbox{min}}}}$ for both types of total errors (Tables \ref{table:3-1} and \ref{table:3-2}). Thus, as shown previously \cite{sdcvao} the parameterizations for the terms of the set $\mathcal{G}_{NN}$ deduced within AQFT demonstrare good agreement with available experimental data. The recent LHC results \cite{ATLAS-NPB-889-486-2014,ATLAS-PLB-761-158-2016,TOTEM-EPJC-76-661-2016} agree well with general trends for global $pp$ scattering parameters (Figs. \ref{fig:1}a, c) especially for $\sigma_{\scriptsize{\mbox{tot}}}^{pp}$ (Fig. \ref{fig:1}a) as well as cosmic ray point at highest available energy $\sqrt{s}=95^{+5}_{-8}$ TeV \cite{TA-PRD-92-032007-2015}. Therefore, one can expect the new data with respect to the DB16 will negligibly affect on the values of fit parameters, especially for high $s_{\scriptsize{\mbox{min}}}$. This qualitative hypothesis is quite confirmed by the fit results for different stages of DB under consideration from Tables \ref{table:3-1} and \ref{table:3-2}. For each of the DB16 and the DB16+ all terms of the set $\mathcal{P}_{NN}$ are the same for Tables \ref{table:3-1} and \ref{table:3-2} within errors for any methods of total uncertainty calculations.

For any DBs considered in the present paper the values of fit parameters do not depend on the method of total error calculations. Thus the parameterizations (\ref{eq:2.1}), (\ref{eq:2.2}) suggested within AQFT shows the strong robustness and stability for values of terms of the set $\mathcal{P}_{NN}$ with regard of different version of modern experimental DB for global scattering parameters $\mathcal{G}_{NN}$. The new data points included in the DB16+ lead only to the slightly poorer fit qualities with regard to the DB16 at corresponding $s_{\scriptsize{\mbox{min}}}$. Despite of the agreement between the values of terms from the set $\mathcal{P}_{NN}$ for DB16 and DB16+ the new data points lead to the some changes in the energy dependence of the global scattering parameters from $\mathcal{G}_{NN}$ due to variations of fit parameters. These changes are in the multi-TeV domain and mostly for $\rho$-parameter. As seen from Figs. \ref{fig:1}c, \ref{fig:2}c the DB16+ provides some vanishing of the spread between fit curves at low and high $s_{\scriptsize{\mbox{min}}}$ and what is more important the fit curve at $\sqrt{s_{\scriptsize{\mbox{min}}}}=30$ GeV changes from the flat (Fig. \ref{fig:1}c) on the decrease (Fig. \ref{fig:2}c) at $\sqrt{s} \gtrsim 10$ TeV. The second case agrees better with a qualitative view of the energy dependence of $\rho$-parameter joined for $pp$ and $\bar{p}p$ at $\sqrt{s} \geq 10$ GeV \cite{TOTEM-EPJC-76-661-2016}. The similar situation is seen for the $\rho^{\bar{p}p}$ at migration from the DB16 (Fig. \ref{fig:1}d) to the DB16+ (Fig. \ref{fig:2}d) with much more marked vanishing of the difference between fit curves at $\sqrt{s_{\scriptsize{\mbox{min}}}}=5$ and 30 GeV. The fit curves at low $\sqrt{s_{\scriptsize{\mbox{min}}}}=3$ and 5 GeV demonstrate the slow growth for $\sigma_{\scriptsize{\mbox{tot}}}^{\bar{p}p}$ (Figs. \ref{fig:1}b, \ref{fig:2}b) with regard of $\sigma_{\scriptsize{\mbox{tot}}}^{pp}$ and opposite behavior for $\rho(\sqrt{s})$ for $pp$ (Figs. \ref{fig:1}c, \ref{fig:2}c) and $\bar{p}p$ (Figs. \ref{fig:1}d, \ref{fig:2}d) scattering in multi-TeV domain. These results, apparently, contradict the general expectations on the basis of Pomeranchuk theorem and make the approximations at higher $\sqrt{s_{\scriptsize{\mbox{min}}}}$ more preferable from the point of view of asymptotic behavior of the energy dependence of global scattering parameters from the set $\mathcal{G}_{NN}$. The new experimental results for $pp$ scattering parameters at RHIC energies as well as for $\rho^{pp}$ and $\bar{p}p$ scattering observables in multi-TeV domain are important to clarify the situation and more unambiguous physical conclusions.

Detailed analysis of the relations between fit parameters in pairs $\alpha^{\pm}$, $\beta^{\pm}$, $\gamma^{\pm}$ and $\delta^{\pm}$ allows the following conclusions. The asymptotic relations $\alpha^{+}=\alpha^{-}$, $\beta^{+}=\beta^{-}$ are valid at $\sqrt{s_{\scriptsize{\mbox{min}}}} \geq 15$ GeV and $\gamma^{+}=\gamma^{-}$ -- at $\sqrt{s_{\scriptsize{\mbox{min}}}} \geq 20$ GeV within mostly 1 standard deviation (s.d.) for DB16 and DB16+ with method (i) for total error calculation (Table \ref{table:3-1}) with exception of the $\gamma^{\pm}$ at $\sqrt{s_{\scriptsize{\mbox{min}}}}=30$ GeV for DB16. The asymptotic constrain $\delta^{+}=\delta^{-}$ is valid at $\sqrt{s_{\scriptsize{\mbox{min}}}} \geq 15$ GeV mostly within 2 s.d. for both DBs with two exceptions, namely, for DB16 at $\sqrt{s_{\scriptsize{\mbox{min}}}}=30$ GeV and for DB16+ at $\sqrt{s_{\scriptsize{\mbox{min}}}}=15$ GeV (Table \ref{table:3-1}). The situation is the same for $\alpha^{\pm}$, $\beta^{\pm}$ at the harder restriction of total errors within method (ii) in Table \ref{table:3-2} with exception only for DB16 at $\sqrt{s_{\scriptsize{\mbox{min}}}}=15$ GeV. In the last case $\alpha^{+}=\alpha^{-}$, $\beta^{+}=\beta^{-}$ within 2 s.d. Moreover equations $\alpha^{+}=\alpha^{-}$, $\beta^{+}=\beta^{-}$ are valid even at $\sqrt{s_{\scriptsize{\mbox{min}}}}=10$ GeV within 2 s.d. for DB16+. The $\gamma^{\pm}$ are equals for each other for both DBs starting at $\sqrt{s_{\scriptsize{\mbox{min}}}}=15-20$ GeV mostly within 2 s.d. The asymptotic relation for $\delta^{\pm}$ is not valid for DB16 at any $\sqrt{s_{\scriptsize{\mbox{min}}}}$ but $\delta^{+}=\delta^{-}$ within 1 (2) s.d. for DB16+ at $\sqrt{s_{\scriptsize{\mbox{min}}}}=20$ (30) GeV.

The comparison the present day DBs with the previous analysis \cite{sdcvao} as well as between DB16 and DB16+ allows the general suggestion that the sequential addition of experimental data at high and ultra-high energies in the fitted sample leads to the some improvement of the validity of the asymptotic constrains for terms of the set $\mathcal{P}_{NN}$ as well as for decrease of the low value $\sqrt{s_{\scriptsize{\mbox{min}}}}$ for such validity. In present analysis, there is a significantly different $\gamma^{\pm}$ parameter from the asymptotic value $\gamma^{\pm}_{\scriptsize{\mbox{a}}}=6$ which can implies no reaching Froissart--Martin bound even at highest measurable cosmic ray energies.

Based on the fit results (Tables \ref{table:3-1}, \ref{table:3-2}) one can obtain phenomenological predictions for terms of the full set of global scattering parameters $\mathcal{G}_{NN}$. These predictions will be within the framework of the Standard Model (SM) physics without any hypothesis for the contributions of the physics beyond the SM. On the other hand, the predictions below are calculated based on the general theorems and limit constrains of the AQFT and, therefore, these results can be useful for search for signatures for any new physics beyond the SM. As seen in Fig. \ref{fig:2} there are detailed measurements for total cross section and $\rho$ parameter for elastic $pp$ (Figs. \ref{fig:2}a, c) and $\bar{p}p$ (Figs. \ref{fig:2}b, d) scattering in energy ranges of the FAIR ($\sqrt{s}=3.0 - 14.7$ GeV) and NICA ($\sqrt{s}=20 - 25$ GeV) projects. Moreover, the most of the present facilities (RHIC, LHC) as well as the international projects for possible future hadron collider (FCC-hh, SPPC, VLHC) focus on investigations with proton beams \cite{FCC-arXiv-1503.09107,SPPC-arXiv-1507.03224,VLHC-Fermilab-TM-2149-2001}.
Thus we estimate the global scattering parameters for elastic $pp$ scattering for RHIC energy domain of $\sqrt{s}=0.20 - 0.51$ TeV because it would be cover the large gap between ISR and the LHC measurements, at energies $\sqrt{s}=2.76, 13 - 42$ TeV of the LHC,
for future projects of proton collider ($\sqrt{s}=40 - 200$ TeV). Estimations are also calculated for ultra-high energy cosmic rays up to $\sqrt{s}=1$ PeV corresponded to the end of cosmic ray spectrum \cite{Greisen-PRL-16-748-1966} as well as for some higher energies.

Phenomenological predictions are derived for $pp$ terms of the set $\mathcal{G}_{NN}$ based on the fit results for more complete data sample DB16+ with method (ii) for total error calculations (Table \ref{table:3-2}) at any $\sqrt{s_{\scriptsize{\mbox{min}}}}$ under consideration. The results agree with each other within errors for any $\sqrt{s_{\scriptsize{\mbox{min}}}}$ at fixed $\sqrt{s}$. But in general uncertainties growth with increase of $\sqrt{s_{\scriptsize{\mbox{min}}}}$ especially for $\rho$ parameter. Detailed analysis shown that prediction errors are reasonable at $\sqrt{s_{\scriptsize{\mbox{min}}}}=15$ GeV. Moreover, it have been shown above the fit quality and behavior of the energy dependence of terms of the $\mathcal{G}_{NN}$ in very high $\sqrt{s}$ domain are acceptable for this $\sqrt{s_{\scriptsize{\mbox{min}}}}$ as well as the validity of asymptotic relations for terms of the $\mathcal{P}_{NN}$. Therefore, Table \ref{table:3-3} demonstrates the estimations for global parameters in elastic $pp$ scattering obtained in the present study at $\sqrt{s_{\scriptsize{\mbox{min}}}}=15$ GeV. Unfortunately the available experimental samples for $\rho$ parameter in $pp$ and $\bar{p}p$ scattering do not allow the predictions within the present model with reasonable accuracy in the TeV-energy domain already.

As expected the results for $\sigma_{\scriptsize{\mbox{tot}}}^{pp}$ from Table \ref{table:3-3} coincide with estimations at the same $\sqrt{s_{\scriptsize{\mbox{min}}}}$ shown elsewhere \cite{sdcvao} within errors and present study allows us to improve the precision of estimations noticeably for most cases regarding of the previous analysis. The comparison between estimations for $\sigma_{\scriptsize{\mbox{tot}}}^{pp}$ at $\sqrt{s}=30$ TeV and  experimental cosmic ray result from Fly's Eye \cite{Baltrusaitis-PRL-52-1380-1984} described in detail in \cite{sdcvao} is also valid for the present paper with some improvement
($\sigma_{\scriptsize{\mbox{tot}}}^{pp}=120 \pm 9$ mbarn, $\rho^{pp}=0.14 \pm 0.14$). As seen from Figs. \ref{fig:1}a, \ref{fig:2}a fits by (\ref{eq:2.1}) within AQFT quite agree with cosmic ray measurement for $\sigma_{\scriptsize{\mbox{tot}}}^{pp}$ at $\sqrt{s}=57 \pm 7$ TeV \cite{PAO-PRL-109-062002-2012} for both DB16 and DB16+ but the agreement is within error bars only at highest measured energy $\sqrt{s}=95^{+5}_{-8}$ TeV \cite{TA-PRD-92-032007-2015}. These qualitative arguments are totally confirmed by calculations, for instance, at $\sqrt{s_{\scriptsize{\mbox{min}}}}=15$ GeV, which provide $\sigma_{\scriptsize{\mbox{tot}}}^{pp}=132 \pm 10$ $(142 \pm 11)$ mbarn, $\rho^{pp}=0.13 \pm 0.14$ $(0.13 \pm 0.15)$ for $\sqrt{s}=57$ (95) TeV. In the first case the value for $\sigma_{\scriptsize{\mbox{tot}}}^{pp}$ agrees quite well with the result based on the analytic amplitude with the Froissart saturation \cite{Block-PRD-84-091501-2011}. But as well as in our previous analysis \cite{sdcvao} the predictions for $pp$ total cross section derived within AQFT in the present paper are significantly smaller than that for energy dependence $\propto \ln^{2}s$ in earlier study \cite{Block-RMP-57-563-1985} while the $\rho^{pp}$ values are in agreement within large uncertainties. The model curves based on the DB16+ with method (ii) for calculation of total errors demonstrate the increase of (absolute) values of global scattering parameters from the set $\mathcal{G}_{NN}$ for $\sqrt{s} > 1$ PeV at any $\sqrt{s_{\scriptsize{\mbox{min}}}}$ with exception of the dependence $\rho^{pp}(\sqrt{s})$ which is almost flat up to much higher energies.

\section{\label{sec4}Final Remarks}
As is well known, the dips in the differential cross section are due to the existence of zeros in the imaginary part of the scattering amplitude. If one assumes that derivative dispersion relations can be applied to small transferred momentum values, then one may associate, at least, the first dip in the differential cross section with the fractal structure in the momentum space. As known from \cite{antoniou_5}, the multiparticle production is controlled by a power-law behavior whose exponent possess a fractal dimension. Therefore, at least in small momentum transfer region, the power may control the total cross section behavior.

In terms of the pomeron exchange, the usual is to consider the exchange of two or three pomerons to tame the total
cross section rise \cite{ddln} considering $s>s_{1}$ here, where $s_{1}$ is some threshold value. If each particle exchanged corresponds to a dip, then one may expect three dips, at least. In terms of fractal dimension, this "partition process" can be viewed as a "fractal dimension partition", i.e.,
the fractal dimension obtained considering $s>s_{1}$ is separated in two or three, each one representing a pomeron exchange. Therefore,
this approach create a multifractal character to the total cross section if $s>s_{1}$.

However, one must take into account the theoretical results concerning the formal fractal theory. The first question is to know if the set with positive fractal dimension in \cite{sdc} or in \cite{antoniou_5} is a homeomorphism to the Cantor set. As is well known, the Cantor set possess fractal dimension $\ln 2/\ln 3\approx 0.630$ and is defined in the interval $[0,1]$. The second question is to know if the iterated  function system formalism can be applied to this problem. If it is possible, then a theorem due to Peres--Shmerkin \cite{fractal_sum} working with iterated function system states that the sum of two fractal dimensions cannot result in the original fractal
dimension, i.e., if one splits the original data set in two, then the sum of the fractal dimensions of each data set is not the original fractal dimension. A possible physical consequence of this mathematical result is the impossibility to describe the whole
elastic scattering using the multiple pomeron exchanges.

It is important to stress the novel interpretation introduced in \cite{bc, sdcvao_2} concerning the total cross
section in terms of fractal dimension and phase transition. The fractal structure present in the total cross section can viewed as
a manifestation of a Berezinskii--Kosterliz--Thouless (BKT)-like phase transition \cite{bere,kt}. The phase transition occurs at
$\sqrt{s_{1}}\approx 15-25$ GeV and represents a region where the fractal dimension is null. Considering $s<s_{1}$ the fractal
dimension is negative, representing the emptiness of the total cross section; if $s>s_{1}$, the fractal dimension is positive
indicating an increasing of the internal structure complexity. Therefore, the odderon is the negative fractal dimension and the
pomeron is the positive one. In terms of phase transition this can be viewed as follows. The fractal dimension presents in momentum
\cite{antoniou_5,antoniou_1} and energy space can be viewed in the impact parameter formalism as a BKT-like phase transition in the sense that the logarithm of the spatial separation of vortex-antivortex pairs is replaced by the logarithm of a probabilistic area in terms of the inelastic overlap function \cite{sdcvao_2}. Therefore, the BKT-like phase transition is the responsible by the change in the total cross section data behavior. In the slope data set, this phase transition can be observed in a possible crossing between $pp$ and $\bar{p}p$ data sets, exactly as occurring to $\sigma_{\scriptsize{\mbox{tot}}}(s)$.

Two consequent stages are considered for modern database for global scattering parameters in elastic $pp$ and $\bar{p}p$ scattering, namely, the latest PDG sample (DB16) and all available experimental results (DB16+). The analytic parameterizations suggested within AQFT provide the quantitative description of energy dependence of global scattering parameters with robust values of fit parameters for any DBs and methods for calculation of total errors under consideration. The fit qualities are reasonable for wide energy range ($\sqrt{s} \geq 3$ GeV) and the quantity is statistically acceptable at low boundary for fit domain $\sqrt{s_{\scriptsize{\mbox{min}}}} \geq 10$ GeV for method (ii). Based on the fit results the estimations are derived for the total cross section and the $\rho$ parameter in elastic $pp$ scattering at different $\sqrt{s}$ up to energy frontier $\sqrt{s}=10$ PeV. These estimations are in a good agreement, in particular, with available cosmic ray data in multi-TeV energy domain. The model proposed in \cite{sdcvao} considers in its formulation some important results of AQFT and presents an interesting characteristic, its sensitivity, to the minimum $\sqrt{s_{\scriptsize{\mbox{min}}}}$. Although based on asymptotic results, the available datasets are not statistically equivalent in terms of distribution over $\sqrt{s}$. The result is an obvious interference of the low energy dataset on the asymptotic results. Therefore, one may expect that asymptotic models using low energy dataset cannot predict the behavior of the elastic scattering amplitude as $s\rightarrow\infty$. The question is to discover the exact starting energy of applicability of each model. On the other hand, regardless of its natural limitation, asymptotic models are important to the understanding of high energy physics.

\section*{Acknowledgments}
The work
of V.A.O. was supported partly by NRNU MEPhI Academic
Excellence Project (contract No 02.a03.21.0005, 27.08.2013). S.D.C is grateful for UFSCar for the financial support.


\clearpage
\begin{table*}
\caption{\label{table:3-1}Parameters for simultaneous fitting by
equations (\ref{eq:2.1}), (\ref{eq:2.2}) of $\mathcal{G}_{NN}(s)$ at different stages of DB: DB16 (first line) and DB16+ (second line). The total errors are calculated by method (i).}
\begin{center}
\begin{tabular}{lccccccc}
\hline \multicolumn{1}{l}{Parameter} &
\multicolumn{7}{c}{$\sqrt{s_{\scriptsize{\mbox{min}}}}$, GeV} \rule{0pt}{10pt}\\
\cline{2-8}
 & 3 & 5 & 10 & 15 & 20 & 25 & 30 \rule{0pt}{10pt}\\
\hline
$k$, mbarn                 & $47 \pm 6$        & $-52 \pm 18$      & $-119 \pm 40$     & $-58 \pm 42$      & $35 \pm 18$       & $475 \pm 347$     & $417 \pm 290$ \rule{0pt}{10pt}\\
                           & $47 \pm 6$        & $-51 \pm 18$      & $-119 \pm 42$     & $-53 \pm 27$      & $34
\pm 11$       & $488 \pm 369$     & $452 \pm 437$ \\
$\alpha^{+} \times 10^{3}$ & $17 \pm 2$ & $14 \pm 4$ & $10 \pm 5$ & $8 \pm 7$ & $20 \pm 5$ & $13 \pm 6$ & $9 \pm 2$ \\
                           & $20 \pm 2$ & $18 \pm 3$ & $15 \pm 5$ & $13 \pm 6$ & $23 \pm 6$ & $15 \pm 4$ & $16 \pm 8$ \\
$\beta^{+} \times 10^{3}$  & $8.7  \pm 0.3$  & $8.3 \pm 0.5$     & $7.7 \pm 0.7$     & $7.4 \pm 1.0$     & $9.4 \pm 0.9$     & $8.1 \pm 1.0$     & $7.6 \pm 0.2$   \\
                           & $9.1  \pm 0.2$  & $8.8 \pm 0.5$   & $8.2 \pm 0.9$     & $7.9 \pm 0.9$     &$9.9
\pm 1.3$     & $8.2 \pm 0.6$     & $8.4 \pm 1.4$   \\
$-\gamma^{+}$              & $5.888 \pm 0.003$ & $5.894 \pm 0.006$ & $5.902 \pm 0.011$ & $5.906 \pm 0.016$ & $5.870 \pm 0.016$ & $5.892 \pm 0.018$ & $5.904 \pm 0.002$ \\
                           & $5.886 \pm 0.002$ & $5.889 \pm 0.006$ & $5.897 \pm 0.014$ & $5.901 \pm 0.015$ &
$5.87 \pm 0.02$   & $5.896 \pm 0.011$ & $5.89 \pm 0.02$ \\
$\delta^{+}$, mbarn        & $51.4 \pm 0.3$    & $50.6 \pm 0.9$    & $49.0 \pm 1.8$    & $49 \pm 3$        & $56 \pm 4$        & $52 \pm 4$        & $48.7 \pm 1.2$  \\
                           & $51.5 \pm 0.3$    & $51.1 \pm 0.9$    & $50 \pm 2$        & $49 \pm 3$        & $57
\pm 5$        & $50 \pm 2$        & $50 \pm 5$  \\
$\alpha^{-} \times 10^{3}$ & $78 \pm 3$ & $52 \pm 5$ & $36 \pm 3$ & $18 \pm 12$ & $25 \pm 10$ & $20 \pm 14$ & $14 \pm 4$ \\
                           & $75 \pm 3$ & $48 \pm 5$ & $33 \pm 2$ & $17 \pm 9$ & $24 \pm 14$ & $22 \pm 16$ & $20 \pm 14$ \\
$\beta^{-} \times 10^{3}$  & $23.1 \pm 0.7$     & $15.7 \pm 1.1$     & $12.1 \pm 0.3$   & $8.9 \pm 1.8$     & $10.1 \pm 1.7$     & $9 \pm 2$     & $8.4 \pm 0.4$  \\
                           & $22.3 \pm 0.6$     & $14.9 \pm 1.0$     & $11.6 \pm 0.3$   & $8.6 \pm 1.3$     &$10 \pm 2$     & $10 \pm 3$     & $9 \pm 2$  \\
$-\gamma^{-}$              & $5.678 \pm 0.005$ & $5.765 \pm 0.011$ & $5.815 \pm 0.002$ & $5.86 \pm 0.02$   & $5.84 \pm 0.02$   & $5.85 \pm 0.03$   & $5.861 \pm 0.009$ \\
                           & $5.682 \pm 0.005$ & $5.771 \pm 0.010$ & $5.819 \pm 0.011$ & $5.867 \pm 0.017$ &
$5.84 \pm 0.04$   & $5.84 \pm 0.05$   & $5.84 \pm 0.04$ \\
$\delta^{-}$, mbarn        & $127.0 \pm 1.6$   & $95 \pm 3$        & $79.3 \pm 1.0$    & $65 \pm 6$        & $74 \pm 7$        & $73 \pm 10$       & $70 \pm 6$ \\
                           & $125.8 \pm 1.5$   & $94 \pm 3$        & $78.2 \pm 0.9$    & $65 \pm 4$        & $73
\pm 12$       & $77 \pm 19$       & $74 \pm 14$ \\
$\chi^{2}/\mbox{n.d.f.}$   & $428/317$         & $188/248$         &$114/164$          & $87.7/133$        & $63.1/98$         & $33.8/69$         & $32.1/63$  \\
                           & $456/322$         & $203/253$         &$127/169$          & $99.8/138$        &
$74.7/103$        & $45.9/74$         & $43.9/68$  \\
\hline
\end{tabular}
\end{center}
\end{table*}

\begin{table*}
\caption{\label{table:3-2}Parameters for simultaneous fitting by
equations (\ref{eq:2.1}), (\ref{eq:2.2}) of $\mathcal{G}_{NN}(s)$ at different stages of DB: DB16 (first line) and DB16+ (second line). The total errors are calculated by method (ii).}
\begin{center}
\begin{tabular}{lccccccc}
\hline \multicolumn{1}{l}{Parameter} &
\multicolumn{7}{c}{$\sqrt{s_{\scriptsize{\mbox{min}}}}$, GeV} \rule{0pt}{10pt}\\
\cline{2-8}
 & 3 & 5 & 10 & 15 & 20 & 25 & 30 \rule{0pt}{10pt}\\
\hline
$k$, mbarn                 & $42 \pm 5$        & $-47 \pm 14$      & $-109 \pm 33$     & $-43 \pm 33$      & $30 \pm 11$       & $485 \pm 291$     & $411 \pm 305$ \rule{0pt}{10pt}\\
                           & $42 \pm 5$        & $-46 \pm 14$      & $-106 \pm 34$     & $-55 \pm 28$      & $60
\pm 30$       & $477 \pm 267$     & $421 \pm 345$ \\
$\alpha^{+} \times 10^{3}$ & $17 \pm 2$ & $14 \pm 3$ & $11 \pm 5$ & $9 \pm 2$ & $18 \pm 4$ & $14 \pm 6$ & $10 \pm 2$ \\
                           & $20.7 \pm 1.4$ & $19 \pm 2$ & $17 \pm 4$ & $13 \pm 3$ & $25 \pm 8$ & $18 \pm 4$ & $16 \pm 7$ \\
$\beta^{+} \times 10^{3}$  & $8.8  \pm 0.3$  & $8.3 \pm 0.5$   & $7.8 \pm 0.7$     & $7.5 \pm 0.2$   & $9.1 \pm 0.8$     & $8.3 \pm 0.9$     & $7.6 \pm 0.3$   \\
                           & $9.2  \pm 0.2$  & $8.9 \pm 0.4$   & $8.6 \pm 0.8$     & $7.8 \pm 0.4$   &$10.2
\pm 1.8$     & $8.7 \pm 0.7$     & $8.4 \pm 1.2$   \\
$-\gamma^{+}$              & $5.888 \pm 0.002$ & $5.894 \pm 0.005$ & $5.901 \pm 0.010$ & $5.906 \pm 0.003$ & $5.876 \pm 0.013$ & $5.891 \pm 0.016$ & $5.907 \pm 0.004$ \\
                           & $5.886 \pm 0.002$ & $5.889 \pm 0.005$ & $5.893 \pm 0.011$ & $5.904 \pm 0.006$ &
$5.86 \pm 0.03$   & $5.889 \pm 0.011$ & $5.90 \pm 0.02$ \\
$\delta^{+}$, mbarn        & $51.4 \pm 0.2$    & $50.6 \pm 0.8$    & $49.5 \pm 1.7$    & $48.7 \pm 0.7$    & $55 \pm 3$        & $52 \pm 3$        & $47.8 \pm 1.3$  \\
                           & $51.5 \pm 0.2$    & $51.1 \pm 0.7$    & $50.6 \pm 1.9$    & $48.8 \pm 1.2$    & $57
\pm 8$        & $51 \pm 2$        & $49 \pm 4$  \\
$\alpha^{-} \times 10^{3}$ & $79 \pm 2$ & $55 \pm 4$ & $39 \pm 3$ & $24 \pm 12$ & $28 \pm 6$ & $21 \pm 9$ & $23 \pm 13$ \\
                           & $77 \pm 2$ & $52 \pm 4$ & $37 \pm 7$ & $20 \pm 6$ & $26 \pm 11$ & $24 \pm 10$ & $21 \pm 13$ \\
$\beta^{-} \times 10^{3}$  & $23.2 \pm 0.6$     & $16.4 \pm 0.9$     & $12.6 \pm 0.4$   & $10 \pm 2$     &$10.7 \pm 1.1$     & $9.4 \pm 1.5$     & $10 \pm 2$  \\
                           & $22.6 \pm 0.6$     & $15.9 \pm 0.8$     & $12.3 \pm 1.4$     & $9.1 \pm 1.0$     &$10 \pm 2$     & $9.9 \pm 1.7$     & $9 \pm 2$  \\
$-\gamma^{-}$              & $5.679 \pm 0.004$ & $5.757 \pm 0.008$ & $5.810 \pm 0.004$ & $5.86 \pm 0.03$   & $5.834 \pm 0.018$ & $5.85 \pm 0.02$   & $5.84 \pm 0.03$ \\
                           & $5.682 \pm 0.004$ & $5.762 \pm 0.008$ & $5.814 \pm 0.018$ & $5.864 \pm 0.013$ &
$5.84 \pm 0.03$   & $5.84 \pm 0.03$   & $5.85 \pm 0.04$ \\
$\delta^{-}$, mbarn        & $125.9 \pm 1.4$   & $97 \pm 3$        & $80.2 \pm 1.3$    & $67 \pm 7$        & $75 \pm 6$        & $73 \pm 7$       & $74 \pm 8$ \\
                           & $124.9 \pm 1.4$   & $96 \pm 2$        & $79 \pm 5$        & $65 \pm 3$        & $73
\pm 9$        & $76 \pm 9$       & $73 \pm 11$ \\
$\chi^{2}/\mbox{n.d.f.}$   & $575/317$         & $268/248$         &$158/164$          & $122/133$         & $81.3/98$         & $46.7/69$         & $43.4/63$  \\
                           & $605/322$         & $286/253$         &$173/169$          & $136/138$         &
$94.4/103$        & $60.1/74$         & $56.9/68$  \\
\hline
\end{tabular}
\end{center}
\end{table*}

\begin{table*}
\caption{\label{table:3-3} Predictions for $pp$ based on the simultaneous fit of $\mathcal{G}_{NN}$ for DB16+ with method (ii) for calculation of total errors.}
\begin{center}
\begin{tabular}{ccccccccc}
\hline \multicolumn{1}{l}{} &
\multicolumn{8}{c}{$\sqrt{s}$, TeV} \rule{0pt}{10pt}\\
\cline{2-9}
\multicolumn{1}{l}{Parameter} &
\multicolumn{3}{c}{RHIC} &
\multicolumn{5}{c}{LHC-I, II, LHC-ultimate} \rule{0pt}{10pt}\\
 & 0.20 & 0.41 & 0.51 & 2.76 & 13 & 14 & 28 & 42 \\
\hline
$\sigma_{\scriptsize{\mbox{tot}}}^{pp}$, mbarn & $51 \pm 3$ & $59 \pm 3$ & $61 \pm 3$ & $82 \pm 5$ & $106 \pm 7$ & $107 \pm 7$ & $119 \pm 9$ & $126 \pm 10$ \rule{0pt}{10pt}\\
$\rho^{pp}$ & $0.13 \pm 0.09$ & $0.14 \pm 0.10$ & $0.14 \pm 0.10$ & $0.14 \pm 0.12$ & $0.14 \pm 0.13$ & $0.14 \pm 0.13$ & $0.14 \pm 0.14$ & $0.14 \pm 0.14$ \\
\hline
\multicolumn{1}{l}{} &
\multicolumn{8}{c}{SPPC, FCC-hh, VLHC-I, II} \rule{0pt}{10pt}\\
 & 40 & 50 & 70.6 & 100 & 125 & 150 & 175 & 200 \\
\hline
$\sigma_{\scriptsize{\mbox{tot}}}^{pp}$, mbarn & $125 \pm 9$ & $130 \pm 10$ & $136 \pm 11$ & $143 \pm 11$ & $147 \pm 12$ & $151 \pm 12$ & $154 \pm 13$ & $156 \pm 13$ \rule{0pt}{10pt}\\
$\rho^{pp}$ & $0.14 \pm 0.14$ & $0.13 \pm 0.14$ & $0.13 \pm 0.14$ & $0.13 \pm 0.15$ & $0.13 \pm 0.15$ & $0.13 \pm 0.15$ & $0.13 \pm 0.15$ & $0.13 \pm 0.15$ \\
\hline
\multicolumn{1}{l}{} &
\multicolumn{6}{c}{ultra-high energy cosmic rays} &
\multicolumn{2}{c}{higher energies} \rule{0pt}{10pt}\\
 & 110 & 170 & 250 & 500 & 750 &$10^{3}$ & $5 \times 10^{3}$ & $10^{4}$ \\
\hline
$\sigma_{\scriptsize{\mbox{tot}}}^{pp}$, mbarn & $144 \pm 12$ & $153 \pm 13$ & $161 \pm 13$ & $175 \pm 15$ & $184 \pm 16$ & $190 \pm 17$ & $225 \pm 22$ & $241 \pm 24$ \rule{0pt}{10pt}\\
$\rho^{pp}$ & $0.13 \pm 0.15$ & $0.13 \pm 0.15$ & $0.13 \pm 0.15$ & $0.14 \pm 0.16$ & $0.14 \pm 0.16$ & $0.14 \pm 0.16$ & $0.14 \pm 0.17$ & $0.14 \pm 0.18$ \\
\hline
\end{tabular}
\end{center}
\end{table*}

\clearpage

\end{document}